\documentclass[utf8]{frontiersinFPHY_FAMS} 
\setcitestyle{square} 

\usepackage{url,hyperref,microtype,subcaption}
\usepackage[onehalfspacing]{setspace}

\def\keyFont{\fontsize{8}{11}\helveticabold }
\def\firstAuthorLast{B\"otzel {et~al.}} 
\def\Authors{Steffen B\"otzel\,$^{1}$, and Ilya M. Eremin\,$^{1,*}$}
\def\Address{$^{1}$Institut f\"ur Theoretische Physik III, Fakult\"at für Physik und Astronomie, Ruhr-Universität Bochum, 44801 Bochum, Germany }
\def\corrAuthor{Corresponding Author}

\def\corrEmail{Ilya.Eremin@rub.de}

\begin{document}
\onecolumn
\firstpage{1}

\title[Feedback of non-local $d_{xy}$ nematicity on the magnetic anisotropy in FeSe]{Feedback of non-local $d_{xy}$ nematicity on the magnetic anisotropy in FeSe} 

\author[\firstAuthorLast ]{\Authors} 
\address{} 
\correspondance{} 

\extraAuth{}

\maketitle

\begin{abstract}

\section{}
We analyze theoretically the magnetic anisotropy in the nematic phase of FeSe by computing the spin and the orbital susceptibilities  from the microscopic
multiorbital model. In particular, we take into account  both the $xz/yz$
and the recently proposed non-local $xy$ nematic ordering and show that the latter one could play a crucial role in reproducing the experimentally-measured temperature dependence of the magnetic anisotropy. This provides a direct fingerprint of the
different nematic scenarios on the magnetic properties of FeSe.
\tiny
 \keyFont{ \section{Keywords:} iron-based superconductors, nematic ordering, magnetic susceptibility} 
\end{abstract}

\section{Introduction}
Iron-based superconductors offer the opportunity to explore the interplay between electronic nematicity, magnetism and superconductivity. While the broad studies on cuprates already provide insights into the competition between magnetism and superconductivity \cite{RevModPhys.70.897,tranquada2014superconductivity}, the role of the still enigmatic nematic state is of particular interest. Since anisotropy arises in crystal structure, orbital and spin degrees of freedom, it is intricate to decipher the underlying mechanism \cite{fernandes2014drives,bohmer2017nematicity}. In most of the iron pnictides the
structural transition precedes or coincides with the magnetic transition at $T_N$, below which long-range antiferromagnetic order sets in \cite{dai2015antiferromagnetic}, supporting the idea that nematicity (spin nematicity) is driven by magnetic interactions \cite{fernandes2012manifestations}. Note that the  spin-nematic scenario can also lead to
an effective orbital ordering once one takes
the orbital content of the spin fluctuations 
within the so-called orbital-selective spin-fluctuation scenario into account \cite{Fanfarillo2015,Fanfarillo2016,Fanfarillo2018}.

Among various iron-based superconductors, FeSe with a simple crystal structure of the stacked FeSe layers has a rather
unique behavior due to the presence of a marked nematic (structural) transition at $T_S = 90$ K and a transition to superconductivity below $T_c = 9$ K, while magnetic order is absent \citep{hsu2008superconductivity,mcqueen2009tetragonal,PhysRevLett.104.087003,Gallais2016,ColdeaWatsonReview2018,shibauchi2020exotic}. Consequently, orbital degrees of freedom have also been proposed as the underlying mechanism for nematic order in FeSe \citep{baek2015orbital,Su2015,Mukherjee2015,Jiang2016,yamakawa2016nematicity,bohmer2015origin,Xing2017}. The small lattice distortion contrasts with strong in-plane anisotropy of resistivity, magnetic susceptibility, electronic structure and orbital and momentum structure of the superconducting gap \cite{margadonna2008crystal,tanatar2016origin,he2018evidence,chen2019anisotropic,lu2021spin,coldea2018key,xu2016highly,sprau2017discovery,rhodes2018scaling,kushnirenko2018three,liu2018orbital,hashimoto2018superconducting}. The phase diagram of FeSe is rich and sensitive to the application of hydrostatic pressure or chemical substitution \citep{huh2021cu,mizuguchi2009substitution,schoop2011pressure,medvedev2009electronic}. Furthermore, various exotic superconducting states have been recently reported in this compound \citep{shibauchi2020exotic}. 

Recent experiments have overcame the intricacies of the formation of orthorombic domains in FeSe by applying uniaxial strain \citep{PhysRevB.90.121111,watson2017electronic,yi2019nematic,huh2020absence,pfau2021quasiparticle}. Using this technique, the in-plane anisotropy of resistivity, uniform magnetic susceptibility and the Knight shift have been found to exhibit the opposite sign of the anisotropy as compared to iron-pnictides \cite{tanatar2016origin,he2018evidence,Andersen2020,PhysRevB.91.155106,chu2010plane}. Moreover, carefully avoiding eddy-current heating, a slight suppression of the Knight shift in the superconducting state has been measured recently, while superconductivity and nematicity seem to coexist \cite{vinograd2021inhomogeneous,li2022se,pustogow2019constraints}. This agrees with direct magnetization measurements \cite{he2018evidence}. 

From the experimental point of view the systematic
investigation of the band-structure of FeSe by means of
ARPES and quantum oscillations revealed a sizeable deformation of the Fermi surface, that can be described by the interplay of the $d_{xz}$, $d_{yz}$ and $d_{xy}$ orbitals, their spin-orbit coupling and the nematic order \cite{Zhang2015,Suzuki2015,Zhang2016,Watson2017}. Concerning the nematic order there is general understanding about the existence of a $xz/yz$ splitting that changes sign in going from the Brillouin-zone center to momenta around ${\bf Q}_X=(\pi,0)$ and ${\bf Q}_Y=(0,\pi)$ points of the 1-Fe-unit cell Brillouin Zone (BZ) (both folded onto the $M$ point of the folded BZ  with 2-Fe ions per unit cell). This can be represented by a nematic order parameter $\Phi^{xz/yz}=\langle d^\dagger_{xz}d_{xz}-d^\dagger_{yz}d_{yz}\rangle$, that is positive around the $\Gamma$ point and is negative around the $M$ point of the BZ. Accounting for this nematic order is straightforward but yields rather controversial electronic structure. In particular, a large electron pocket with mixed $xz$ and $xy$ character is expected at the $M$ point of the BZ, which has not been resolved in the most recent ARPES measurements in detwinned samples \cite{watson2017electronic,yi2019nematic,huh2020absence}. One approach to explain these experimental data and to suppress modelled contributions associated with this pocket, is to include orbital-selective quasiparticle weights \cite{chen2019anisotropic,Andersen2020,kreisel2017orbital,hu2018orbital}. More recently an alternative scenario has been
proposed \cite{rhodes2021non,islam2021specific,marciani2022resistivity,rhodes2022fese}, where an additional non-local nematic order parameter accounting for the splitting of the $xy$ occupancy in the two electron pockets $\Phi^{xy}=\langle d^{X\dagger}_{xy}d^{X}_{xy}-d^{Y\dagger}_{xy}d^{Y}_{xy}\rangle$  \cite{rhodes2021non,islam2021specific,marciani2022resistivity} plays a crucial role. An important consequence of this scenario is a resulting occurrence of a Lifshitz transition at the $M$ point leaving only one electron pocket at the Fermi level. 

This possibility to have a non-local nematic ordering of the d$_{xy}$ states was previously outlined in the literature \cite{Fernandes2014,Classen2017,Christansen2020}  but assumed to be small in the models having on-site interaction terms only. However, an inclusion of the nearest-neighbor interaction terms (such as nearest neighbor exchange or Coulomb interaction) would change this picture. Note, recent NMR \cite{Liu2020} and ARPES \cite{yi2019nematic,huh2020absence} studies have also suggested that the d$_{xy}$ orbital may be strongly affected by the onset of the nematic state.

In this manuscript we investigate the consequences of the non-local $d_{xy}$-nematic scenario for the magnetic susceptibility taking into account spin and orbital contributions. We analyze its temperature and doping dependencies and compare the results to the available experiments. We show that the non-local nematicity is responsible for the non-monotonic temperature dependence of the susceptibility and predict how it evolves with doping in FeSe$_{1-x}$S$_x$. Finally we studied how  the magnetic anisotropy is affected in the superconducting state.

\section{Methods}
We adopt the low energy model for FeSe, previously employed in  Ref. \cite{rhodes2021non}, and fitted to the available ARPES experiments \cite{watson2017electronic,PhysRevB.91.155106,watson2015emergence}. It is based on the generalized low-energy effective model for iron-based superconductors, formulated by Cvetkovic and Vafek \cite{Vafekmodel}. 

\subsection{Tetragonal state}
In particular, to describe the tetragonal state of FeSe, we start by describing the low-energy states near the corresponding symmetry points of the BZ. In particular, near the $\Gamma$ point of the BZ there are two Fermi surface pockets formed by the hybridized $xz$ and $yz$ orbitals. For completeness we also take the $xy$ orbital band into account which is located approximately 50meV below the Fermi level \cite{watson2015emergence}. Its inclusion allows to treat correctly the spin-orbit coupling (SOC) within the $t_{2g}$ manifold. The states can be described by the spinor $\Psi^\dagger_{\Gamma,\mathbf{k}} = (d^\dagger_{xz\uparrow},d^\dagger_{yz\uparrow},d^\dagger_{xy\downarrow})$, where the other ('lower') spin part of the Hamiltonian is related by symmetry. For readability we omit the momentum index in the creation and annihilation operators here and in what follows.
In particular, for each momentum $\mathbf{k}$ the Hamiltonian is given by
\begin{equation}
	H^{\Gamma}(\mathbf{k}) = 
	\begin{pmatrix}
	\epsilon_{xz,\mathbf{k}} - \frac{b}{2} (k_x^2-k_y^2)& b k_x k_y - i\frac{\lambda_1}{2} & \frac{i\lambda_2}{2} \\
	b k_x k_y + i\frac{\lambda_1}{2} & \epsilon_{yz,\mathbf{k}} - \frac{b}{2} (k_y^2-k_x^2)& -\frac{\lambda_2}{2} \\
	-\frac{i\lambda_2}{2} & -\frac{\lambda_2}{2} & \epsilon_{xy,\mathbf{k}}
	\end{pmatrix},
\end{equation}
where $\epsilon_{\mu,\mathbf{k}} = \epsilon_{\mu,0} - \frac{\mathbf{k}^2}{2m_{\mu}} - \mu$.  Figure \ref{Figure1}A shows the resulting band dispersion for the tetragonal state. The orbital weights of the bands are illustrated in the usual red-green-blue-color scheme for $d_{xz}$, $d_{yz}$ and $d_{xy}$-orbitals, respectively. 

To describe the electronic states near the $M$-point of the 2-Fe unit cell we introduce the spinors $\Psi^\dagger_{X/Y,\mathbf{k}} = (d^\dagger_{yz/xz\uparrow},d^\dagger_{xy\uparrow})$. The momenta are defined with respect to $X/Y$ point of the 1-Fe unit cell, which are folded into the $M$-point. The Hamiltonian reads
\begin{equation}
H^{X/Y}(\mathbf{k}) =
\begin{pmatrix}
A^{X/Y}_{yz/xz,\mathbf{k}} & -iV^{X/Y}_{\mathbf{k}}, \\
iV^{X/Y}_{\mathbf{k}} & A^{X/Y}_{xy,\mathbf{k}}. \\
\end{pmatrix}.
\end{equation}
and the elements are 
\begin{equation}
A^{X/Y}_{\mu,\mathbf{k}} = \epsilon_{\mu,0} - \frac{\mathbf{k}^2}{2m_{\mu}} \mp \frac{a_{\mu}}{2}(k_x^2-k_y^2) - \mu,
\end{equation}
\begin{equation}
V^{X/Y} = \sqrt{2}v k_{y/x} + \frac{p_1}{\sqrt{2}}(k_{y/x}^3 + 3 k_{y/x} k_{x/y}^2) \mp  \frac{p_2}{\sqrt{2}} k_{y/x} (k_{x}^2 - k_{y}^2 ).
\end{equation}
The orbitals are coupled by the in-plane SOC $\lambda_2$
\begin{equation}
	H^{M}_{\text{SOC}} = \frac{\lambda_2}{2} \sum_{\mathbf{k}} 
	\left(
        i^{X\dagger}_{yz\uparrow}    d^Y_{xy\downarrow}
	    + d^{X,\dagger}_{xy\uparrow}  d^Y_{xz\downarrow}
	+h.c. \right)
	.
\end{equation}
The band dispersion for the tetragonal state near the $M$ point is shown in the bottom panel of the figure \ref{Figure1}D. 
Note the value of SOC is taken different if the orbitals originate from the same sublattice in the 1-Fe unit cell, $\lambda_1 = 23$ or from the two different sublattices,  $\lambda_2 = 4$ meV \cite{scherer2017interplay}. This agrees with the experimental ARPES observation, which measured different values of the SOC near different symmetry points of the BZ \cite{Borisenko2015,Day2018}.

\subsection{Nematic state}

To describe the nematic state in FeSe, which forms below the $T_S = 90$ K, we add the nematic order to the Hamiltonian in a phenomenological fashion. We assumed the nematic order parameters follow a mean-field temperature dependence $\Phi(T) = \Phi(0)\sqrt{1 - T/T_S}$. As described in the Introduction we distinguish two scenarios: \textbf{Scenario A} containing an order parameter $\Phi^{h,e}$ that lifts the  $d_{xz}$ and $d_{yz}$ degeneracy, and \textbf{Scenario B} in which an additional non-local $d_{xy}$-order parameter $\Phi_{xy}$ and a $d_{xy}$-Hatree shift $\Delta\epsilon_{xy}$ are added \cite{rhodes2021non}.

\begin{figure}[h!]
	\begin{center}
		\includegraphics[width=17.5cm]{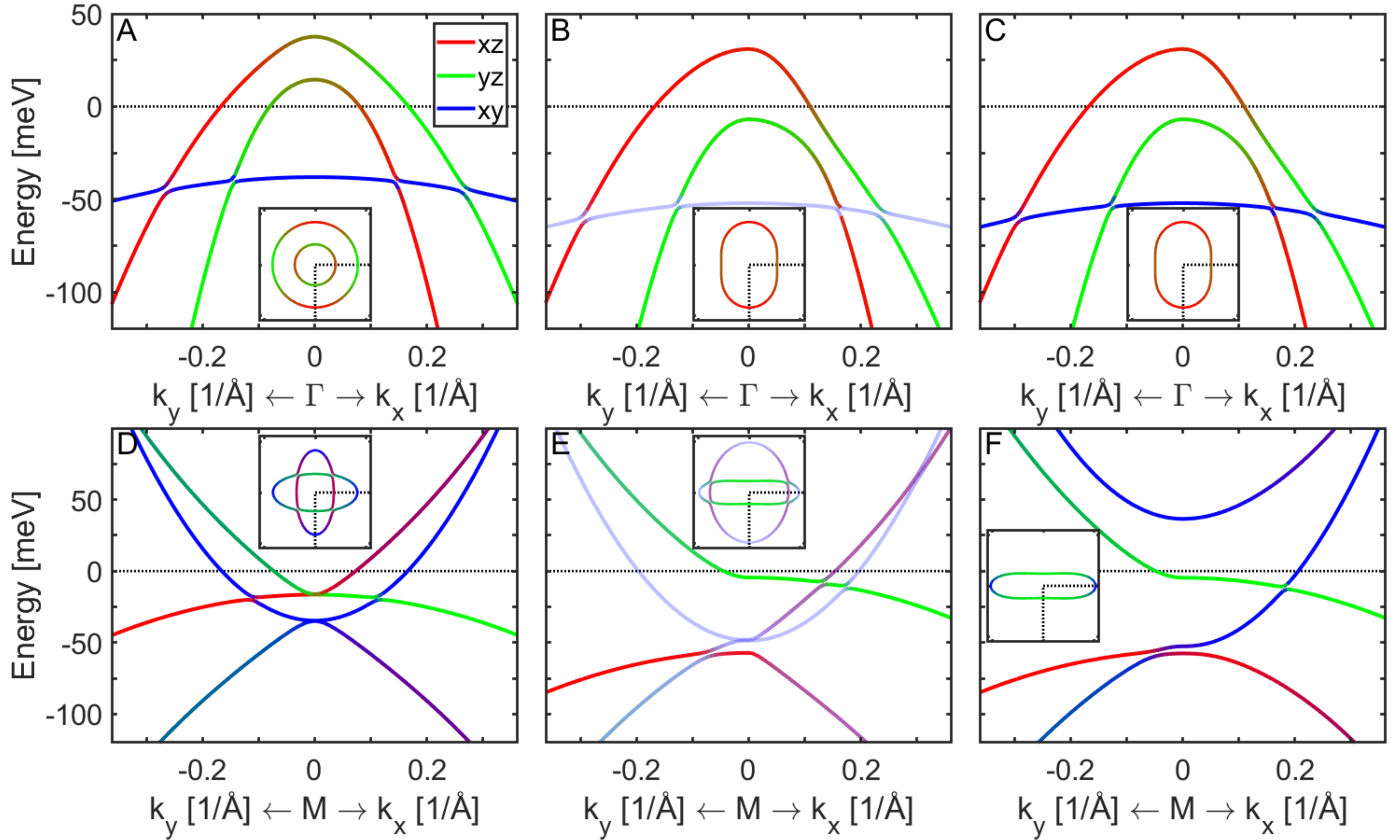}
	\end{center}
	\caption{The band structure of the 2-Fe Brillouin zone around the $\Gamma$ and the $M$ point is presented comparing the tetragonal state and scenarios A and B for the nematic state in FeSe as described in the text.
	Panels \textbf{(A)-(C)} show the energy dispersion near the $\Gamma$ point in the tetragonal state \textbf{(A)} and two nematic states without (Scenario A) \textbf{(B)} and with \textbf{(C)} non-local $d_{xy}$ nematicity (Scenario B). Panels \textbf{(D)-(F)} display the corresponding band structure near the $M$ point. Red, green and blue colors illustrate the orbital weights of the $d_{xz},d_{yz}$ and $d_{xy}$ Fe-orbitals, respectively. The insets show the corresponding Fermi surface and demonstrate the cuts performed. Shading according to the $d_{xy}$ orbital weight in panels \textbf{(B)} and \textbf{(E)} reflects the $Z$-factors used in Ref. \cite{Andersen2020}.} \label{Figure1}
\end{figure}

In particular, the $xz/yz$ nematic order parameter near the $\Gamma$ point can be described as: \begin{equation}
	H^{\Gamma}_{\text{nem}} = \Phi^h \sum_{\mathbf{k},\sigma \in (\uparrow,\downarrow)} \left( d^\dagger_{xz,\sigma} d_{xz,\sigma} - d^\dagger_{yz,\sigma} d_{yz,\sigma,} \right). 
\end{equation} 
For a given value of the SOC $\lambda_2$, adding a nematic order parameter yields a Lifshitz transition such that one of the Fermi surface pockets near the $\Gamma$ point sinks below the Fermi level as shown in panels \ref{Figure1}B and \ref{Figure1}C for $\Phi^h(10 \text{ K}) = 15$ meV. 
Near the $M$ point the $xz/yz$ nematic order has opposite sign to $\Phi^h$ and we also include an additional non-local $d_{xy}$ nematic order \cite{rhodes2021non}:
\begin{equation}
	H^{M}_{\text{nem}} = \sum_{\mathbf{k},\sigma} \left(  \Phi^e (d^{Y\dagger}_{xz,\sigma} d^{Y}_{xz,\sigma} - d^{X\dagger}_{yz,\sigma} d^{X}_{yz,\sigma})
	+ (\Delta\epsilon_{xy} + \Phi_{xy}) d^{Y\dagger}_{xy,\sigma} d^{Y}_{xy,\sigma}
	+ (\Delta\epsilon_{xy} - \Phi_{xy}) d^{X\dagger}_{xy,\sigma} d^{X}_{xy,\sigma}	
 \right). 
\end{equation}

In the conventional scenario of nematicity (Scenario A), only $\Phi^e(10 \text{ K}) = -26$ meV is included and the resulting band structure near the $M$ point is illustrated in Figure \ref{Figure1}E. Observe two electron pockets at the $M$ point, which exist for all temperatures. To connect Scenario A with experimental ARPES results in the nematic state, which do not observe the larger electron pocket, orbital-selective quasiparticle weights are introduced in various works \cite{chen2019anisotropic,sprau2017discovery,Andersen2020,kreisel2017orbital, Kostin2018}, see also for a  recent review Ref. \cite{Kreisel2020}. Within this scenario the quasiparticle weight of the $d_{xy}$ orbital is claimed to be much smaller than that of the $d_{yz}/d_{xz}$ orbitals. This would lead to a suppression of the quasiparticle weight of the $d_{xy}$ dominated bands and especially of the outer electron pocket at $M$ point, making $d_{xy}$ orbital states invisible in the ARPES experiment. This is illustrated in panels \ref{Figure1}B and \ref{Figure1}E by orbital-selective shading. The idea of orbital-selective spectral weights is motivated also by the previous results of the dynamical mean-field theory calculations \cite{Yin2011}. This scenario was claimed to be in agreement with STM measurements of the electronic structure \cite{Kostin2018}, the superconducting gap properties \cite{sprau2017discovery}, and the magnetic susceptibility measured by inelastic neutron scattering \cite{chen2019anisotropic}.  A problem of this scenario is that both in the tetragonal and the nematic
state of FeSe, bands of $d_{xy}$ orbital character have been identified around the $M$ point \cite{coldea2018key}. Although $d_{xy}$ orbital appears indeed to exhibit a larger effective mass
renormalization than the other two orbitals, it is not sufficiently large to mask $d_{xy}$ spectral weight completely in the ARPES experiments \cite{watson2015emergence}.

Within the Scenario B one introduces the non-local nematic order within the $xy$-orbital: $\Phi_{xy}(10 \text{ K}) = 45$ meV and $\Delta\epsilon_{xy}(10 \text{ K}) = 40$ meV  and the resulting band structure is shown in Figure \ref{Figure1}F. As the band structure evolves from the tetragonal to the nematic state, an additional Lifshitz transition occurs around 70 K, leaving a single electron Fermi surface pocket at 10 K. The Scenario B assumes that the nematic ordering near the $M$ point does not cause a minor perturbation of the electronic structure, but can lift an entire electron pocket away from the Fermi level. Its consequences for various experiments were reviewed in Ref. \cite{rhodes2022fese} and present an alternative description of the nematicity in FeSe. The chemical potential $\mu$ is renormalized to fulfill the Luttinger theorem and is set to zero at 10 K \citep{rhodes2021non}.

\subsection{Superconducting state}

It was argued in the past that within both the Scenario A and the Scenario B the superconducting order parameter and its angular dependence on the Fermi surface pockets can be equally well described within a microscopic description \cite{kreisel2017orbital,rhodes2021non}. This indicates that superconductivity just adopts the existing electronic structure in the nematic state without further significant feedback on the nematicity. So within Scenario B we use the microscopic description of Ref. \cite{rhodes2021non}. In particular, below the superconducting transition at $T_c = 8$ K, we additionally include superconducting gaps in the $d_{xz}$ and $d_{yz}$ orbitals and employ previously found values from a microscopic calculations describing an extended s-wave gap structure \citep{rhodes2021non}. The values are obtained solving the gap equations at $T=0$ K focusing on the bands crossing the Fermi surface. Possible $d_{xy}$ contributions to the pairing interaction are neglected. To describe the temperature dependence we assume a typical BCS form $\Delta_{\mu}(T) = \Delta_{\mu}(0 \text{ K}) \tanh{( 1.74\sqrt{T_c/T - 1 } )}$. For the $\Gamma$ point we use
\begin{equation}
	H^{\Gamma}_{\text{sc}} = \sum_{\mathbf{k},\mu \in \{\text{xz,yz}\}} \left( \Delta^{\Gamma}_{\mu} d^\dagger_{\mu,\uparrow,\mathbf{k}} d^\dagger_{\mu,\downarrow,-\mathbf{k}} + h.c. \right)
\end{equation} 
with $\Delta^\Gamma_{xz}(0 \text{ K}) = -0.1$ meV and $\Delta^\Gamma_{yz}(0 \text{ K}) = -5.3$ meV. For the $M$ point we use
\begin{equation}
	H^{M}_{\text{sc}} = \sum_{\mathbf{k},\mu \in \{\text{yz(X),xz(Y)}\}} \left( \Delta^{X/Y}_{\mu} d^\dagger_{\mu,\uparrow,\mathbf{k}} d^\dagger_{\mu,\downarrow,-\mathbf{k}} +h.c. \right)
\end{equation} 
with $\Delta^Y_{xz}(0 \text{ K}) = 3.1$ meV and $\Delta^X_{yz}(0 \text{ K}) = 1.6$ meV. The resulting Bogoliubov-de Gennes Hamiltonian is diagonalized numerically.
    
\subsection{Magnetic susceptibility}
Within random phase approximation (RPA) analysis of the magnetic susceptibility $\chi$ \citep{maier2009neutron,graser2009near,kemper2010sensitivity} the bare part of the susceptibility is given by a combination of normal and anomalous contributions ($GG$ and $FF$-terms)
\begin{equation}
	\label{Eq:general_susceptibility}
	\chi_{\eta_1\eta_4}^{\eta_2\eta_3}(q) =   \sum_{k}
	\left[F_{\eta_2\eta_4}(k)\bar{F}_{\eta_1\eta_3}(k+q) - G_{\eta_2\eta_1}(k) G_{\eta_4\eta_3}(k+q) \right], 
\end{equation}
where $G$ and $F$ denote the single-particle normal and anomalous Green's functions and the $FF$-term vanishes above $T_c$. We use the short hand notations $k = (\mathbf{k},i\omega_n)$ and $\eta = (\mu,\sigma)$ denoting orbital and spin degrees of freedom. The sum of Matsubara frequencies is carried out analytically and yields a Lindhard-type expression for the bare susceptibilities. Further details can be found in the Supplementary Information.

Within RPA we include the local interactions from a Hubbard-Hund Hamiltonian
\begin{align}
	H_{int} &= 
	\frac{U}{2}\sum_{\mu\sigma\mathbf{k}\mathbf{k'}\mathbf{q}} 
	d^{\dagger}_{\mu\sigma\mathbf{k+q}}
	d^{\dagger}_{\mu\bar{\sigma}\mathbf{k'-q}}
	d_{\mu\bar{\sigma}\mathbf{k'}}
	d_{\mu{\sigma}\mathbf{k}}
	+ \frac{U'}{2}\sum_{\mu\neq\nu\sigma\sigma'\mathbf{k}\mathbf{k'}\mathbf{q}} 
	d^{\dagger}_{\mu\sigma\mathbf{k+q}}
	d^{\dagger}_{\nu\sigma'\mathbf{k'-q}}
	d_{\nu{\sigma'}\mathbf{k'}}
	d_{\mu{\sigma}\mathbf{k}} \nonumber \\
	&+ \frac{J}{2}\sum_{\mu\neq\nu\sigma\sigma'\mathbf{k}\mathbf{k'}\mathbf{q}} 
	d^{\dagger}_{\nu\sigma\mathbf{k+q}}
	d^{\dagger}_{\mu\sigma'\mathbf{k'-q}}
	d_{\nu\sigma'\mathbf{k'}}
	d_{\mu{\sigma}\mathbf{k}}
	+ \frac{J'}{2}\sum_{\mu\neq\nu\sigma\mathbf{k}\mathbf{k'}\mathbf{q}} 
	d^{\dagger}_{\mu\sigma\mathbf{k+q}}
	d^{\dagger}_{\mu\bar{\sigma}\mathbf{k'-q}}
	d_{\nu\bar{\sigma}\mathbf{k'}}
	d_{\nu{\sigma}\mathbf{k}}.
	\label{Eq:HubbardHundHam}
\end{align}	
In what follows we set the intraorbital Coulomb repulsion interaction $U = 1$ eV. For the interorbital Coulomb repulsion $U'$, the residual Hund interaction $J$ and the pair hopping $J'$ terms we employ the standard spin-rotational relations $U' = U -2J$ and $J=J'$ and set $J/U = 1/6$ similar to previous studies \cite{chen2019anisotropic,Andersen2020}. 

The RPA treatment yields a Dyson type equation with a symbolic solution
 \begin{equation}
 	\label{Eq:RPA_solution}
 	\chi_{RPA}  = \left[ 1 - \chi_0 \hat{U} \right]^{-1} \chi_0.
 \end{equation} 
The interaction matrix $\hat{U}$ results from equation \ref{Eq:HubbardHundHam} and the free spin and orbital indices are contracted with matrix elements from the magnetic operator, which is a combination of spin and angular momentum operators 
\begin{equation}
	\mu^a_{\eta_1\eta_4} = \left(  \frac{g}{2}  {\sigma}^a_{\sigma_1 \sigma_4} \delta_{\mu_1\mu_4} + L^a_{\mu_1\mu_4} \delta_{\sigma_1 \sigma_4} \right).
\end{equation} 
The magnetic susceptibility tensor component $\chi^{ab}$ results from the contraction of the external spin and orbital indices of the solution with the $a$-th and $b$-th component of the  operator. We can separate the spin $\chi_{\text{spin}}$, the orbital $\chi_{\text{orb}}$ and mixed $\chi_{\text{mix}}$ components of the susceptibility. In this manuscript we focus on the static limit of uniform susceptibility $\chi^{ab}(q\rightarrow0)$ and further details of the calculation can be found in the supplementary.   

\section{Results}
In the following we present the results for the magnetic susceptibility $\chi(q\rightarrow0)$ and its anisotropy following Scenario B. Note that within Scenario A the behavior of the magnetic susceptibility was considered in Ref. \cite{Andersen2020}. One of the main findings of this study was that in order to reproduce the correct sign of the magnetic anisotropy at low temperatures, {\it i.e.}
$\chi^{yy} < \chi^{xx}$, the quasiparticle weight of various orbitals has to fulfill  $Z_{xy}<Z_{xz}<Z_{yz}$.
Within Scenario B this assumption is not necessary as the $d_{xy}$ orbital is weakly present at the Fermi level and this should yield in principle the right sign of the magnetic anisotropy. We show in the Supplementary that further factors can affect the signs of the anisotropy as well and the situation depends on the details of the electronic structure in both Scenarios.  
In addition, in our case we find the dominance of the spin component of the susceptibility $\chi_{\text{mix}} < \chi_{\text{orb}} < \chi_{\text{spin}}$, which is a consequence of the low-energy model we used. Moreover, we find that RPA primarily amplifies the spin contribution. Including the whole 3$d$ state manifold would increase the temperature-independent orbital contribution to the susceptibility, which can be treated as a constant. 

Nevertheless, one of the interesting questions is how to distinguish between both scenarios and whether one is able to identify an additional Lifshitz transition associated with the removal of the second electron pocket from the Fermi level at the $M$ point both theoretically and experimentally. To do so we calculate the temperature dependence of the static uniform susceptibility and compare our results with experimental data of Ref. \cite{he2018evidence} where the standard notations $x{\parallel}a$, $y{\parallel} b$ and $z{\parallel}c$ are used. 

\begin{figure}[h!]
	\begin{center}
		\includegraphics[width=17.5cm]{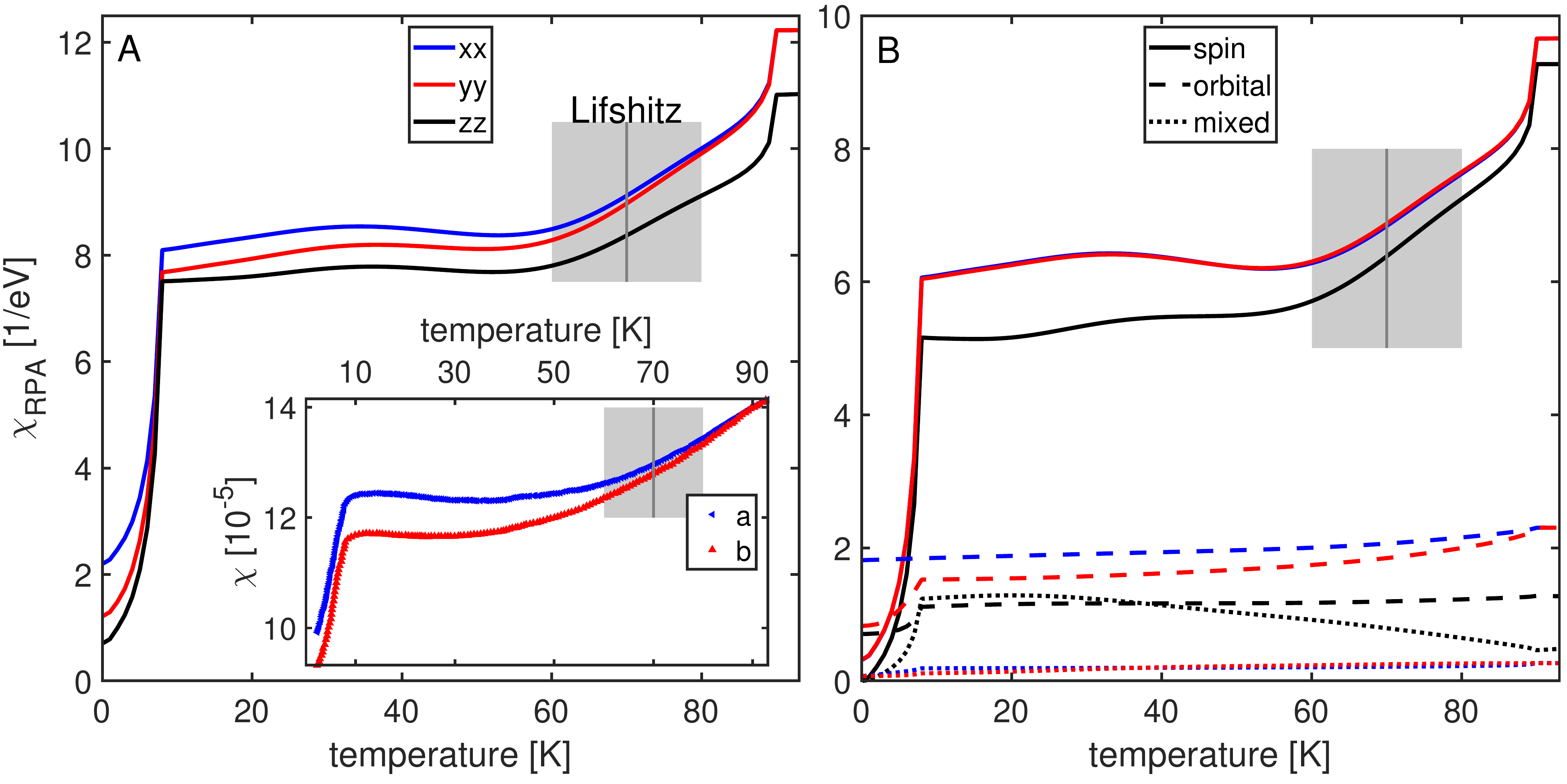}
	\end{center}
	\caption{Calculated temperature dependence of the magnetic susceptibility $\chi(q\rightarrow0)$ calculated with an RPA approach. In \textbf{(A)} the total diagonal components $\chi^{xx},\chi^{yy}$ and $\chi^{zz}$ of the susceptibility calculated for scenario B are presented as blue, red and black curves, respectively. The corresponding spin, orbital and mixed contributions to the different components are shown in \textbf{(B)} as solid, dashed and dotted curves, respectively. The inset in \textbf{(A)} shows the experimental data extracted from \citep{he2018evidence} for comparison. The markers do not represent data points from the experiment. The gray boxes illustrate the decrease in susceptibility due to the Lifshitz transition of the $Y$-band. The boxes are centered around the Lifshitz transition temperature (dark gray line). Notice that the blue solid curve for the $\chi^{xx}_{\text{spin}}$ component in  \textbf{(B)} is almost completely hidden behind the red curve for the $\chi^{yy}_{\text{spin}}$ component due to the tiny in-plane anisotropy in the spin susceptibility.} \label{Figure2}
\end{figure}

The temperature dependence of the static uniform susceptibility is shown in Figure \ref{Figure2}A.  For comparison we also display the experimental data from Ref. \cite{he2018evidence} in the inset. Above the nematic transition the magnetic anisotropy is governed by the spin-orbit coupling and shows an easy-plane anisotropy $\chi^{xx}=\chi^{yy}>\chi^{zz}$. Below $T_S$, one observes two characteristic features. First we find that $\chi^{xx}$ becomes progressively larger than $\chi^{yy}$ as a result of the nematic order and the splitting between them increases continuously upon decreasing temperature. At the same time all three components of the magnetic susceptibility show a non-monotonic temperature dependence characterized by a rapid decrease below $T_S$ and a plateau-like behavior, which starts around 60 K. Both, a continuous increase of the anisotropy upon lowering temperature and the non-monotonic temperature dependence of the susceptibilities agree well with the available experimental data of Ref. \cite{li2020spin}, shown in the inset. 

To understand better the origin of these effects we show in Figure \ref{Figure2}B the spin, the orbital, and the mixed susceptibilities. Note, the spin component of the magnetic susceptibility is the largest in magnitude and its non-monotonic temperature dependence is connected to the Lifshitz transition within Scenario B. In this scenario, the non-local $d_{xy}$ nematic order parameter induces the shift of the larger electron pocket away from the Fermi level, which occurs in the temperature interval $60$ K $< T <T_S$ and is clearly visible in the spin part of each ($xx$, $yy$ and $zz$) component of the magnetic susceptibility. At the same time the main origin of the continuous increase of the magnetic in-plane anisotropy  $\chi^{xx} > \chi^{yy}$ comes from the orbital part of the magnetic susceptibility as evident from Figure \ref{Figure2}B. The in-plane anisotropy for the spin susceptibility is three orders of magnitude smaller within our modeling, which is a consequence of the small in-plane SOC $\lambda_2$ present at the $M$ point and of the out-of-plane SOC between $xz$ and $yz$ orbitals present at hole pockets only transferring out-of-plane anisotropy to the spin susceptibility. Overall the orbital susceptibility is less sensitive to the orbital content of the Fermi surface but to overall orbital structure at low energies. This indicates that not only the pockets at the Fermi level but the overall electronic structure (also away from the Fermi level) is responsible for the continuous increase of the magnetic anisotropy and its correct sign. Note that within both Scenario A and Scenario B of the nematicity in FeSe the correct sign of the magnetic in-plane anisotropy in the uniform susceptibility can be successfully reproduced. What, however, remains unclear in the Scenario A is whether the non-monotonous temperature dependence, which in Scenario B is a clear signature of a Lifshitz transition, could be reproduced in the Scenario A as well. 

We also note that in the iron pnictides, where nematicity coexists with the antiferromagnetic order an in-plane anisotropy of the magnetic susceptibility would give a corresponding feedback on the magnetic order parameter and also enhance the anisotropy of the spin susceptibility below the magnetic transition temperature, see for example Ref. \cite{He2017} . However, as the magnetic transition in FeSe appears only upon applying pressure and bearing in mind that the in-plane anisotropy we found is only few percent of the total magnitude of the spin susceptibility, the anisotropy of the orbital susceptibility will be the main origin of the magnetic anisotropy in FeSe.

To complete the analysis we also computed the change of the uniform superconductivity in the superconducting state. Due to the spin singlet superconducting order parameter in FeSe we see a sharp drop in the uniform susceptibility upon entering the superconducting state, see Figure \ref{Figure2}A. This is due to the strong reduction of the spin component of the susceptibility, as evident from Figure \ref{Figure2}B. Note the small admixture of the spin-triplet component of the superconducting gap due to the finite spin-orbit coupling does not affect much the magnetic anisotropy in the spin part. The residual contribution to the magnetic susceptibility stems largely from the orbital part of the susceptibility and corresponds to inter-orbital contributions, which is slightly affected by the onset of superconductivity. The decrease in $\chi^{yy}_{\text{orb}}$ yields an increase of the in-plane anisotropy and results from intra-band contributions, which appear due to hybridization of $d_{yz}$ and $d_{xy}$ orbitals at Fermi level (compare Figure \ref{Figure1}F).

\begin{figure}[h!]
	\begin{center}
		\includegraphics[width=17.5cm]{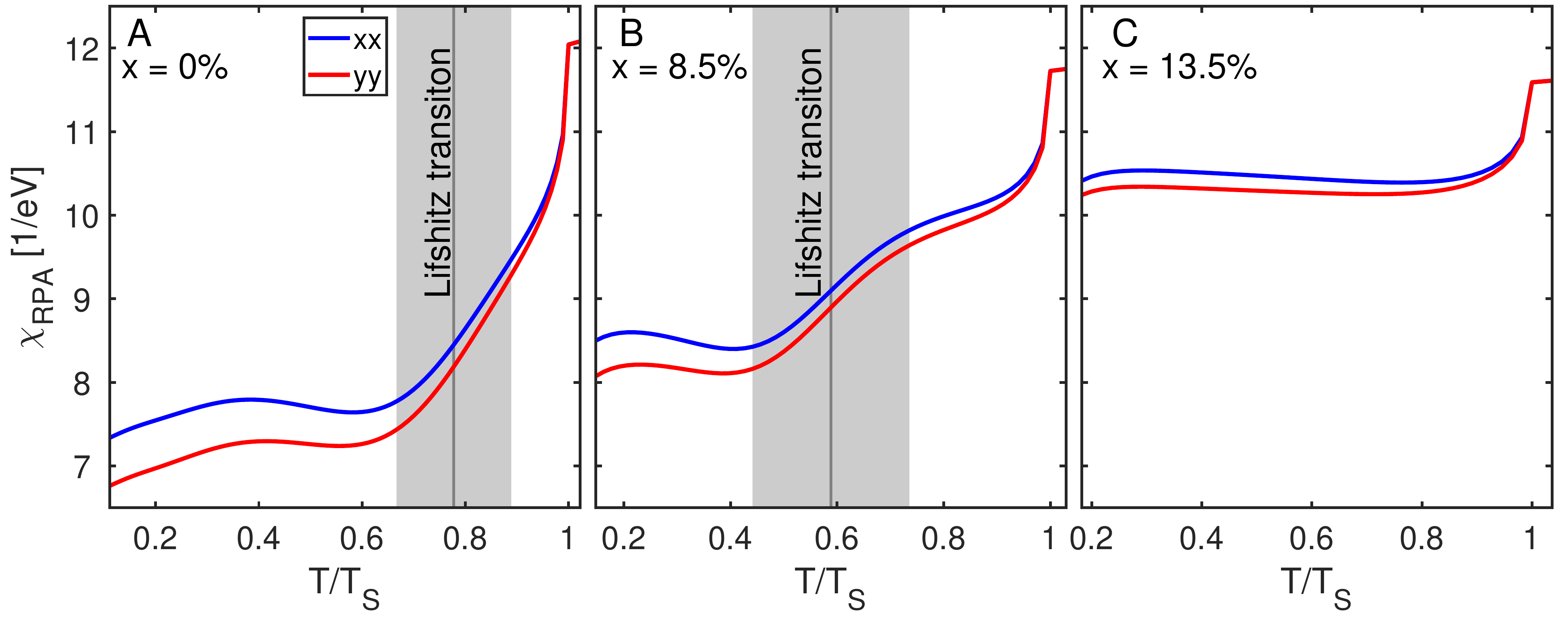}
	\end{center}
	\caption{Calculated effect of sulphur doping of FeSe$_{1-x}$S$_{x}$ on temperature dependence of the magnetic susceptibility $\chi(q\rightarrow0)$. The panels \textbf{(A)}-\textbf{(C)} compare the doping values $x = 0$, $x = 0.085$ and $x = 0.135$. As in Figure \ref{Figure2}, the gray boxes illustrate the Lifshitz transition region of the $Y$-band. For $x = 13.5$, the Lifshitz transition is absent.
	} \label{Figure3}
\end{figure}

To make a qualitative prediction on the evolution of the uniform susceptibility and its in-plane magnetic anisotropy in Figure \ref{Figure3}, we present our calculations for the susceptibility in the doped  FeSe$_{1-x}$S$_{x}$ compounds, following previous analysis of superconductivity and nematicity \cite{rhodes2021non}. In particular, we show $\chi^{xx}$ and $\chi^{yy}$ for $x=0$, $x= 0.085$ and $x = 0.135$ in panels \ref{Figure3}A-\ref{Figure3}C, respectively. The doping evolution of the nematic order parameters is modeled with a mean-field dependence $\Phi(x) = \Phi(0)\sqrt{1 - x/x_0}$. Here we estimate $x_0 = 0.18$, $T_S(x = 0.085) =68$ K and $T_S(x = 0.135) = 55$ K  from interpolation of the phase diagram given in \cite{coldea2019evolution}. The Lifshitz transition is shifted to lower (relative) temperatures with increasing doping and is expected to disappear at around $x \approx 0.13$. Thus, for intermediate doping, we expect to see further the Lifshitz transition and significant changed temperature dependence for doping above $x = 0.135$. As the orbital part is not as sensitive to the Lifshitz transition, significant residual in-plane anisotropy is expected for all dopings.    

\section{Discussion and Conclusion}
Our calculated results for the uniform magnetic susceptibility within the non-local $d_{xy}$ nematicity agree well with experimental observations including the Knight shift measurements \cite{baek2015orbital,bohmer2015origin,he2018evidence,vinograd2021inhomogeneous,li2022se,li2020spin}. Within this scenario there is no necessity to employ orbital-selective quasiparticle weights. Furthermore, this scenario can also successfully explain the non-monotonic temperature dependence of the uniform susceptibility as resulting from the Lifshitz transition of the larger electron pocket, which rapidly shifts away from the Fermi level within $60$ K $< T <T_S$
temperature interval.

Our studies suggest that both spin and orbital contributions and their temperature dependencies are important in describing the uniform magnetic susceptibility or the Knight shift. While the spin part provides the main contribution to the temperature dependence, the orbital part is crucial for a sizeable in-plane anisotropy as observed in experiments. The temperature dependence of the spin part fits very well to the scenario of a Lifshitz transition at $Y$ point, which appears due to the inclusion of the non-local $d_{xy}$ nematicity. 
The finite temperature dependence of the orbital susceptibility breaks linear relation between the bulk magnetic susceptibility and the Knight shift and naturally explains the observed Knight shift anomaly \cite{vinograd2021inhomogeneous,li2020spin}. We note further that the orbital contribution can be in fact larger if one includes larger energy window for the considered model. This should result in the larger in-plane anisotropy but will not affect the non-monotonic temperature dependence of the spin susceptibility.

Our results for the superconducting state also agree qualitatively with recent measurements. The decrease below $T_c$ is seen in measurement of the static, uniform bulk susceptibility  \cite{he2018evidence,li2020spin} and also in the Knight shift measurements \cite{vinograd2021inhomogeneous,li2022se}. Moreover, in agreement with our results, a slight enhancement of in-plane anisotropy is observed \cite{vinograd2021inhomogeneous}. 

In summary, we have studied the uniform magnetic  susceptibility for a model of FeSe and FeSe$_{1-x}$S$_x$ compounds with particular attention on the consequences of non-local $d_{xy}$-nematicity. We associate the corresponding Lifshitz transition of a $Y$ electron pocket with a sharp decrease of the spin component of the magnetic susceptibility, whereas large in-plane anisotropy of the magnetic susceptibility is linked to the orbital susceptibility. The hierarchy of the anisotropy depend on orbital structure of the electronic bands at and away from the Fermi level where the orbital selective quasiparticle weights could be only one potential factor affecting the anisotropy.

We further note that the nematicity also affects the anisotropy of the spin fluctuations at the antiferromagnetic momentum near $(\pi,0)$ or $(0,\pi)$ wavevectors. This anisotropy we also found in our calculations by computing the spin response at large momenta (not shown). This behavior is quite similar in both scenarios of nematicity, while we expect a different temperature dependence of the uniform magnetic susceptibility at $q=0$ in both cases.

\section*{Conflict of Interest Statement}

The authors declare that the research was conducted in the absence of any commercial or financial relationships that could be construed as a potential conflict of interest.




\section*{Acknowledgments}
We are thankful to Jakob B\"oker and Luke Rhodes for fruitful discussions. This work was supported by a joint NSFC-DFG grant (ER 463/14-1). We also acknowledge support by the Open Access Publication Funds of the Ruhr-Universit\"at Bochum. 

\section*{Supplemental Data}
The Supplementary is available on request.

\section*{Data Availability Statement}
The code generated during the current study is available on reasonable request.

\bibliographystyle{Frontiers-Vancouver} 
\bibliography{test}


\section*{Figures}




\end{document}